# Development and Test of a Large-aperture Nb$_3$Sn Cos-theta Dipole Coil with Stress Management

I. Novitski, A.V. Zlobin, M. Baldini, S. Krave, D. Orris, D. Turrioni, and E. Barzi, *Senior Member, IEEE*

*Abstract*—Large-aperture high-field superconducting (SC) magnets are used in various accelerator systems of particle accelerators/colliders. Large Lorentz forces and mechanical stresses can degrade or damage the brittle SC coils. The stress-managed cos-theta (SMCT) coil is a new concept for high-field and/or large-aperture accelerator magnets based on low-temperature and high-temperature superconductors. This concept was proposed and is being developed at Fermilab in the framework of the US Magnet Development Program (US-MDP). The SMCT structure is used to reduce the large coil deformations under Lorentz forces and, thus, the excessively large strains and stresses in the coil. A 120-mm aperture two-layer Nb$_3$Sn SMCT dipole coil has been developed at Fermilab to demonstrate the SM concept including coil design, fabrication technology and performance. The first SMCT demo coil was assembled with a 60-mm aperture Nb$_3$Sn coil inside a dipole mirror configuration to be tested separately and in series with the insert coil. This paper summarizes the SMCT coil design and parameters, the mirror magnet assembly, and the results of separate SMCT coil test in the dipole mirror configuration.

*Index Terms*— Accelerator magnet, dipole mirror, magnetic field, mechanical structure, Rutherford cable, stress management.

## I. Introduction

An innovative stress-management (SM) concept for cos-theta (CT) coils (SMCT coil concept) has been proposed and is being developed at Fermilab [1], [2]. A large-aperture Nb$_3$Sn SMCT dipole coil was designed and manufactured to validate and test the SM concept including coil design, fabrication technology, and performance. The SMCT coil structure has a complex 3D geometry which is difficult to make using conventional machining. To overcome this problem, advanced Additive Manufacturing (AM) technologies based on the Direct Metal Laser Melting with stainless steel powder and CNC post-processing were used to produce accurate metallic SMCT coil parts [3]. The first SMCT coil (SMCT1) was fabricated and assembled with a 60-mm aperture Nb$_3$Sn coil, originally developed for MDPCT1 dipole [4], inside a dipole mirror magnet.

SMCT1 coil tests are to be performed in a dipole mirror structure in two different configurations - SMCTM1a with powered 2-layer SMCT1 coil and SMCTM1b with the SMCT coil connected in series with a 60-mm aperture 2-layer dipole coil. The test goals are to prove the SMCT coil concept in 2-layer and 4-layer mirror configurations; demonstrate that the magnet can reach the targeted quench current at the established pre-load; study magnet training, training memory after thermal cycle, ramp rate and temperature dependences of the magnet quench current; and test the SMCT1 coil quench protection parameters.

This paper summarizes the SMCT1 coil design and parameters, the coil main fabrication steps and instrumentation, and its assembly in the dipole mirror structure. The results of the SMCT1 coil separate test are also presented and discussed.

## II. SMCT Coil Design and Technology

### A. SMCT Coil Design

A 3D view of the large-aperture SMCT dipole coil with transverse and longitudinal cuts at the coil non-lead end is shown in Fig. 1. The SMCT1 coil consists of 2-layers. The turns in each layer are combined in 5 blocks wound into a stainless-steel structure with 5 mm radial and azimuthal block separation [5]. To produce a dipole field in the magnet aperture, the number of turns in the blocks approximately follows the cos-theta distribution. The number of turns in the inner-layer (IL) and outer-layer (OL) coil blocks are summarized in Table I.

TABLE I
DESIGN NUMBER OF TURNS IN COIL BLOCKS.

| Block number* | 1 | 2 | 3 | 4 | 5 |
|---|---|---|---|---|---|
| Inner layer | 12 | 11 | 10 | 7 | 4** |
| Outer layer | 15 | 14 | 13 | 10 | 6** |

\* counted from the mid-plane
\*\* during coil winding the number of turns in both pole blocks was reduced to 3 and 5 respectively due to the limited space

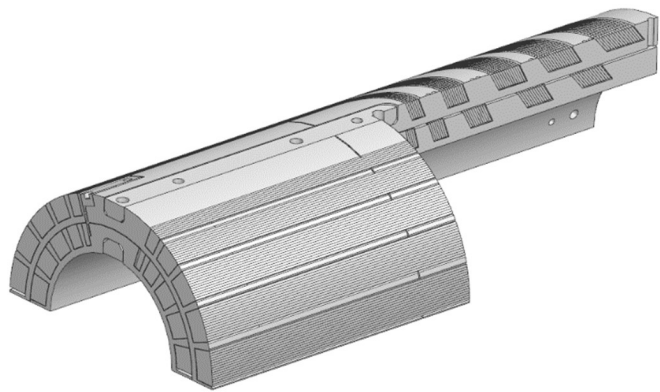

Fig. 1. 3D view of the large-aperture two-layer SMCT coil.

This work is supported by Fermi Research Alliance, LLC, under contract No. DE-AC02-07CH11359 with the U.S. Department of Energy
Authors are with the Fermi National Accelerator Laboratory (FNAL), Batavia, IL 80510 USA (e-mail: zlobin@fnal.gov).

The coil uses a 15.1 mm wide and 1.319 mm (average) thick 40-strand Rutherford keystoned cable with a keystone angle of 0.805 degree. The cable is made of 0.7 mm diameter $Nb_3Sn$ composite wire with a Cu/nonCu ratio of 1.13 and $J_c$ at 15 T and 4.2 K of 1500 A/mm$^2$. This cable was developed and fabricated at Fermilab [6] and used in 2-layer coils developed for the MBHSP dipoles for the LHC upgrades [7] and outer coils in the 4-layer MDPCT1 [8] dipole.

Each coil block in its specific groove in both layers is surrounded with 0.5 mm thick ground insulation made of mica and S2-glass blanket. Consequently, azimuthal and radial components of the Lorentz force in each block are not accumulated but transmitted to the coil and magnet structures partially bypassing the coil blocks. The coil inner diameter is 123 mm, leaving ~0.5 mm of radial space for the coil radial insulation of the inner coil. The coil outer diameter is 206 mm.

### B. SMCT1 Coil Fabrication and Instrumentation

The details of the SMCT1 coil fabrication are reported in [5]. Each coil layer was wound into the SMCT1 coil structure using non-reacted $Nb_3Sn$ cable (React-&-Wind approach) insulated with 0.075 mm thick E-glass tape wrapped with ~45% overlap. The azimuthal and radial dimensions of the SMCT grooves were slightly oversized to provide room for the block insulation and for the $Nb_3Sn$ cable transverse expansion after reaction. The gaps between coil end and central blocks enabled the cable axial extension throughout coil heating and reaction. The coil was reacted in Argon in a 160-hour 3-step cycle with $T_{max}$=658°C for 48 hours. After reaction, flexible Nb-Ti cables were spliced to the $Nb_3Sn$ coil leads and covered by lead end (LE) saddle extension blocks made of stainless-steel. The coil was wrapped with a 0.125 mm thick S2-glass cloth, potted with CTD101K epoxy resin, and cured at 125°C for 16 hours.

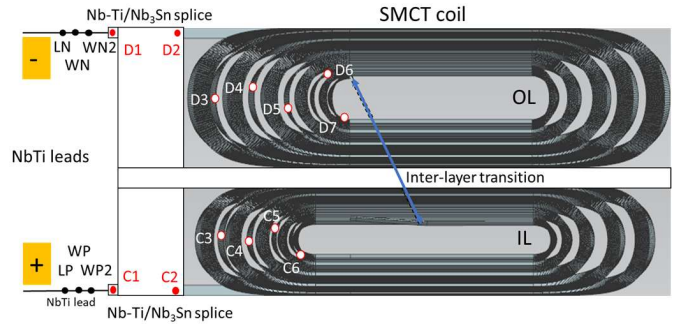

Fig. 2. SMCT1 coil schematic with Nb-Ti/Nb$_3$Sn lead splices, flexible Nb-Ti leads and voltage tap positions.

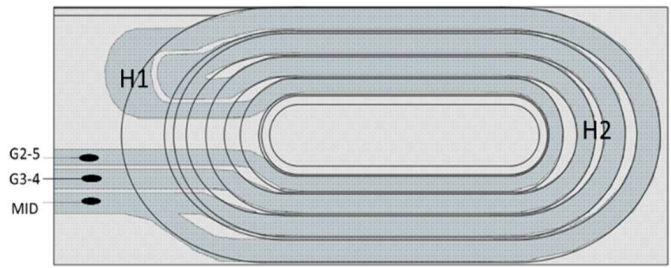

Fig. 3. A layout of quench protection heaters on the SMCT1 coil outer layer. Protection heater H1 covers the midplane block 1 and block 4 and is connected to the Heater Firing Unit (HFU) by MID and G2-5 leads. Protection heater H2 covers blocks 2 and 4 and is connected to the HFU by MID and G3-4 leads.

Coil electrical connections and voltage tap (VT) positions are shown in Fig. 2. VTs were installed on the cable in the inter-block transition areas to detect quench in each block. VTs on the SMCT coil inner layer are marked as C3-C6 and VTs on the coil outer layer are marked as D3-D7. Unfortunately, during magnet assembly VTs C3-C6 lost contact to the coil.

### C. Mirror Magnet Assembly, Coil Preload, Instrumentation

The SMCT1 coil will be tested in a 4-layer dipole mirror configuration using the MDPCT1 mechanical structure and two-layer inner coil [9]. The inner MDP and outer SMCT1 coils are separated by three layers of Kapton of total thickness of 0.3 mm. Two quench protection heaters, made of 0.025 mm thick stainless-steel strips etched on the 0.050 mm thick Kapton sheet (substrate), cover the four largest outer blocks of the SMCT1 coil (Fig. 3). The heaters were placed between the first and the second layer of the four-layer ground insulation. Each layer thickness is 0.125 mm. The total Kapton thickness between the heater and the coil is 0.175 mm. There are no quench protection heaters on the inner coil.

The coil assembly, surrounded by a 1-mm thick 316L stainless steel shell, was placed inside the bottom part of the horizontally-split iron yoke. Analysis shows [5] that the mirror with horizontally split yoke provides lower stresses in the inner coil during magnet assembly and operation making it more suitable. The yoke is made of AISI 1020 iron laminations with an outer diameter of 587 mm, connected by strong 7075-T6 aluminum I-clamps, and enclosed in a 12.5-mm thick 316 stainless-steel skin. To accommodate the SMCT coil with larger OD, the inner diameter of iron laminations was increased from 196.1 mm to 209.5 mm.

The 3D view of the four-layer coil assembly in a dipole mirror structure is shown in Fig. 4. Coil ends are supported by two independent systems. The outer SMCT1 coil is supported by eight 24.5-mm diameter rods and 50 mm thick end plates. The inner coil is supported by four 30-mm diameter rods and 50 mm thick inside and 30 mm thick outside end plates (see Fig. 4).

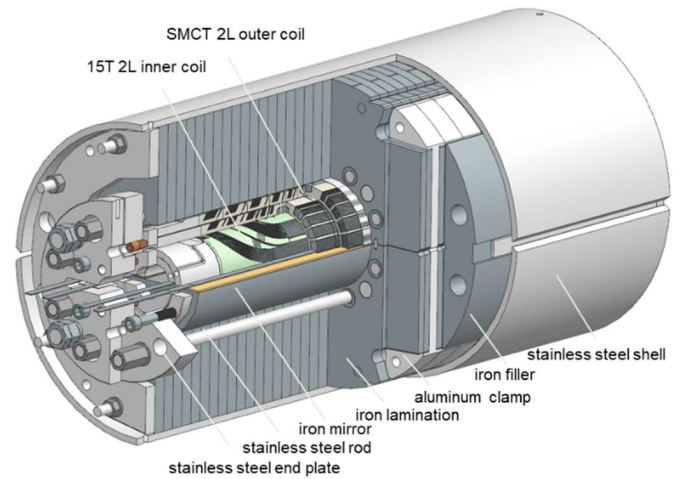

Fig. 4. 4-layer mirror magnet design concept.

The coil preload during assembly is produced by a combination of coil mid-plane, intercoil and coil-yoke radial shims; yoke-clamp interference; yoke-skin shims; and skin tension after welding. During magnet cool-down to the operation temperatures, the coil stress is managed by the gap between the yoke blocks. The details of the mechanical analysis and material properties are reported in [5]. After assembly the maximum stress in the inner MBH coil is less than 130 MPa and in the SMCT1 coil it is less than 50 MPa. After cooling-down the maximum stress in the inner MBH coil increases to 173 MPa and in the SMCT1 coil to ~80 MPa.

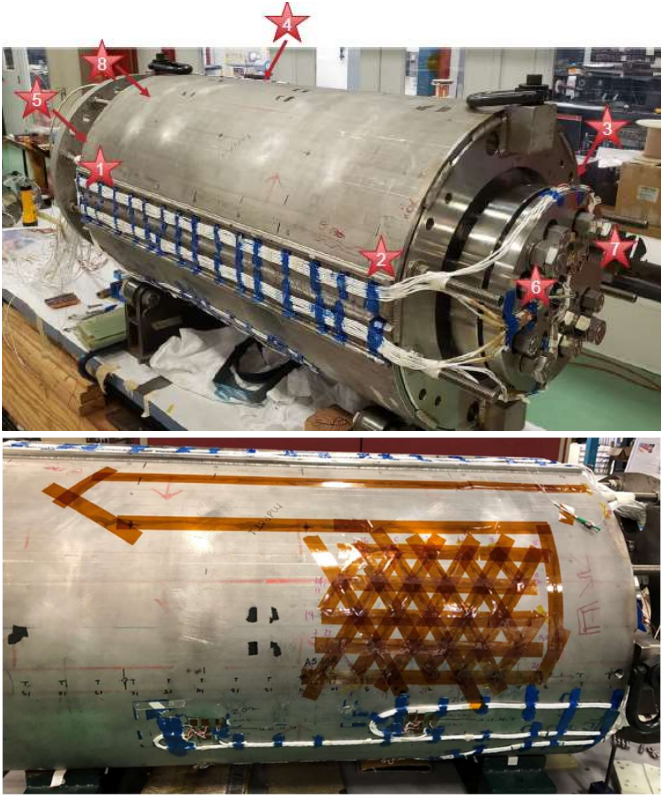

Fig. 5. Position of acoustic sensors (top) and fiber optic sensors (bottom) in the SMCTM1 cold mass.

### D. Instrumentation

SMCTM1 instrumentation includes voltage taps for quench studies and splice resistance measurements; strain gauges on coils, poles, bullets/rods and shell for strain monitoring during cool down, warm up and quenches; protection heaters for quench protection of the superconducting coil. In addition, acoustic sensors were used on both magnet ends for independent quench characterization. Fiber optics sensors were installed on the magnet shell for strain measurements. The position of acoustic and fiber optic sensors on the SMCTM1 cold mass is shown in Fig. 5. Temperature sensors on both magnet ends provide monitoring the magnet/coil temperatures during the tests.

### E. Test Configurations and Conductor Limit

The mirror test is to be done in two configurations. First, the SMCT1 coil will be powered independently (SMCTM1a). Then it will be connected in series and powered with the inner coil (SMCTM1b). In SMCTM1a configuration, the SMCT1 coil is tested with respect to the azimuthal Lorentz forces. In SMCTM1b configuration, the radial Lorentz force from the inner coil will be added.

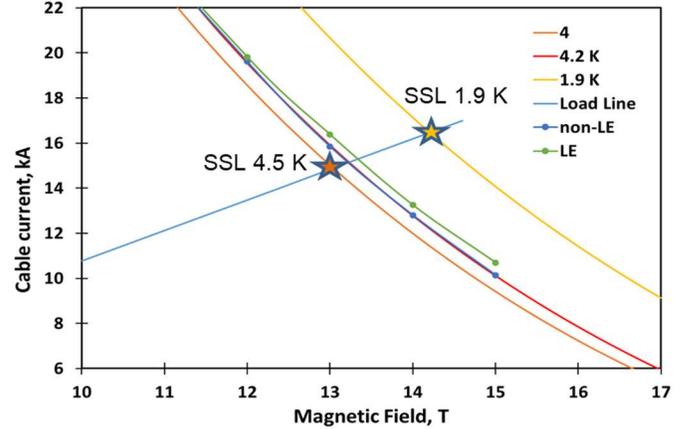

Fig. 6. $I_c(B)$ curves of the 40-strand Nb$_3$Sn cable based on witness sample data for the SMCT1 coil measured at 4.2 K and scaled to 4.5 and 1.9 K and the corresponding 2D coil load line in the dipole mirror test configuration.

The $I_c(B)$ curves of the 40-strand Nb$_3$Sn cable based on witness sample data for the SMCT1 coil and the corresponding coil load line (calculated in a 2D approximation) of the SMCT1 coil in dipole mirror test configuration are shown in Fig. 6. The calculated conductor limits at 1.9 K and 4.5 K of the individually powered SMCT1 coil are 14.2 T and 13.1 T reached at 16.5 kA and 14.8 kA respectively.

### F. SMCTM1a Test

SMCTM1a test started in September of 2023 and then continued in November after a thermal cycle (TC). The test started with magnet training (step I) followed by quench current ramp rate studies (step II) at 1.9 K and temperature dependence measurement (step III) in the temperature range of 1.9-4.5 K. After TC the magnet training memory, the ramp rate and temperature dependences (steps IV-VI) were measured at 1.9 K and 4.5 K. The sequence of magnet quenches and test steps are shown in Fig. 7.

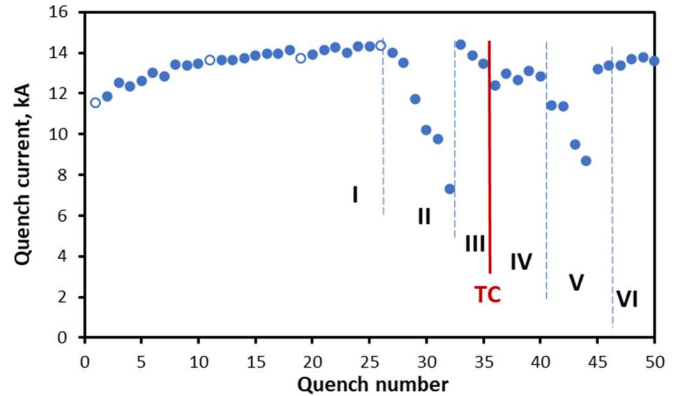

Fig. 7. Magnet test steps and quench sequence.

Magnet training at 1.9 K before and after TC is shown in Fig. 8. SMCT1 coil training started at 11.521 kA which corresponds to 70% of the coil short sample limit at 1.9 K. In 18 quenches the magnet current reached 86% of the SSL and the maximum field in the SMCT1 coil of 12.5 T. All the quenches in SMCT1 coil were detected in the coil inner layer except the 1st quench that started in the middle block 3, and 11th, 19th and 26th quenches which started in the pole block of the coil outer layer. Due to loss of VTs on the inner-layer coil blocks the precise location of the quench origin in the inner layer was not possible. The first quench after TC was 12.408 kA which is only slightly higher than the first training quench. Thus, the magnet showed noticeable loss of the training memory with all the quenches in the coil inner layer. Due to slow training rate, magnet training was stopped after 4 training quenches.

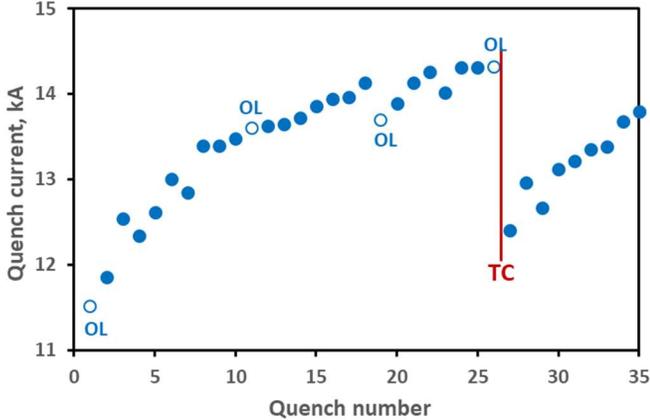

Fig 8. Quench current vs. quench number at 20 A/s. The OL quenches are shown by open markers. Magnet training before and after TC is not completed.

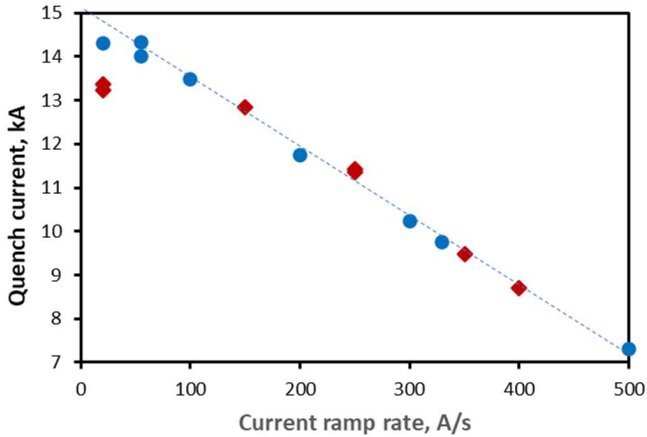

Fig 9. Quench current vs current ramp rate at 1.9 K measured before (circles) and after (dimonds) TC. Magnet training at 20 A/s is not completed.

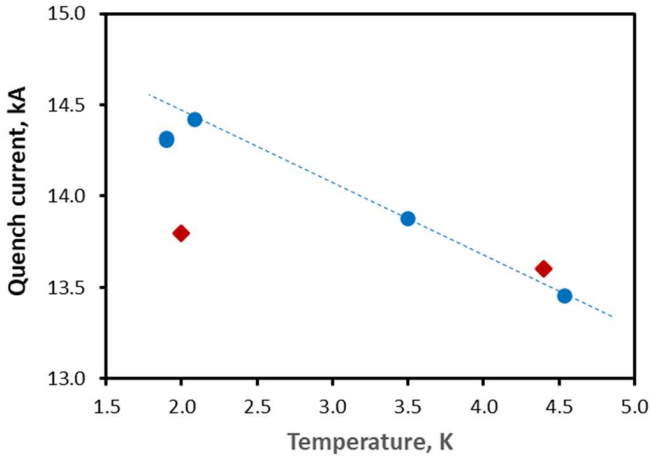

Fig 10. Quench current vs. temperature measured before (circles) and after (dimonds) TC. Magnet training at ~2 K before and after TC is not completed.

Due to the magnet long training before and after TC at 1.9 K, the coil degradation can be estimated using the ramp rate and the temperature dependences of the coil quench current, shown in Figs. 9 and 10 and measured before and after TC.

The ramp rate dependence of the SMCT1 coil quench current shows a practically linear reduction of the coil quench current with increasing current ramp rate. Such dependence indicates that the AC losses in the cable are dominated by the hysteresis loss in the superconductor. The other two AC loss components, i.e., eddy current losses in strands and in cable, are relatively small. Linear extrapolation of the coil quench current to zero ramp rate gives a value of the maximum quench current of 15.0 kA. It corresponds to a maximum field in the coil of 13.1 T which is 91% of the expected SSL at 1.9 K.

The temperature dependence of the SMCT1 coil quench current measured before TC shows that the maximum field in the coil linearly reduced from 12.7 to 12 T which corresponds to 90% of the conductor limit at 4.5 K. The ~9% reduction of magnet maximum achieved quench current with respect to the SSL at both 1.9 K and 4.5 K could be due to conductor degradation during coil fabrication and mirror assembly. Due to lack of voltage taps in the coil inner layer, the precise location of the degradation region is not possible.

Consistency of ramp rate quenches at current ram rates above 150 A/s and quench current at ~4.5 K before and after TC confirms that the coil has not degraded after TC.

Measured values of splice resistances are 0.34 and 0.30 nOhm for the inner-layer and outer-layer splices respectively. The RRR measurements and the information from the strain and acoustic gauges, and the fiber optics will be reported separately.

## III. Conclusion

The SMCT coil concept was proposed and is being studied at Fermilab for high-field and/or large-aperture accelerator magnets based on low-temperature and high-temperature superconductors. The SMCT coil structure allows reducing coil deformations under Lorentz forces and, thus, the high stresses in the coil, dangerous for coils made of brittle superconductors.

A 120 mm aperture $Nb_3Sn$ SMCT1 dipole coil was designed and built at Fermilab to validate and study the SM coil concept. The SMCT1 coil was tested in dipole mirror configuration. In the first test, after a relatively short training, the SMCTM1a mirror magnet with the SMCT1 coil powered individually, has reached a maximum field in the coil of 12.7 T at 1.9 K and 12.0 T at 4.5 K which corresponds to ~90% of its SSL. After TC the magnet showed significant loss of its training memory. However, no conductor degradation was found after TC.

SMCT1 coil test in a 4-layer mirror configuration will follow.

### Acknowledgments

The authors thank V.V. Kashikhin for the coil magnetic optimization; J. Carmichael and J. Coghill for the end part design; A. Rusy and J. Karambis for the technical support of this work; G. Chlachidze, S. Stoynev and technical staff of Fermilab's Magnet Test Facility for the help with magnet testing.

## References

[1] V.V. Kashikhin, I. Novitski, A.V. Zlobin, "Design studies and optimization of a high-field dipole for a future Very High Energy *pp* Collider", *Proc. of IPAC2017*, Copenhagen, Denmark, May 2017, p. 3597.

[2] A.V. Zlobin et al., "Conceptual design of a 17 T Nb$_3$Sn accelerator dipole magnet," *Proc. of IPAC2018*, WEPML027, 2018, p. 2742.

[3] I Novitski et al., "Using Additive Manufacturing technologies in high-field accelerator magnet coils," *CEC-ICMC'21*, FERMILAB-CONF-21-369-TD, 2021.

[4] I. Novitski et al., "Development of a 15 T Nb$_3$Sn Accelerator Dipole Demonstrator at Fermilab," *IEEE Trans. on Appl. Supercond.*, vol. 26, no. 4, Jun. 2016, Art. no. 4001007.

[5] I. Novitski et al., "Development of 120-mm diameter Nb$_3$Sn dipole coil with stress management", *IEEE Trans. on Appl. Supercond.*, vol. 32, no. 6, Sept. 2022, Art. no. 4006005.

[6] E. Barzi et al., "Development and Fabrication of Nb$_3$Sn Rutherford Cable for the 11 T DS Dipole Demonstration Model", *IEEE Trans. on Appl. Supercond.*, vol. 22, no. 3, June 2012, Art. no. 6000805.

[7] A. V. Zlobin et al., "Development and test of a single-aperture 11T Nb$_3$Sn demonstrator dipole for LHC upgrades", *IEEE Trans. on Appl. Supercond.*, vol. 23, no. 3, June 2013, Art. no. 4000904.

[8] A.V. Zlobin et al., "Development and First Test of the 15 T Nb$_3$Sn Dipole Demonstrator MDPCT1", *IEEE Trans. on Appl. Supercond.*, vol. 30, no. 4, 2020, 10.1109/TASC.2020.2967686.

[9] A.V. Zlobin et al., "Reassembly and Test of High-Field Nb$_3$Sn Dipole Demonstrator MDPCT1", *IEEE Trans. on Appl. Supercond.*, vol. 31, no. 5, August 2021, 10.1109/TASC.2020.2967686.